\documentclass{emulateapj}

\usepackage{graphicx}
    \usepackage{epsfig}
    \usepackage{natbib}
    \usepackage{xcolor}
    \usepackage{float} % control places of eps files
    \usepackage{amsmath}
    \usepackage{lizz}

\shorttitle{Buried NLS1} \shortauthors{Pan et al.}

\slugcomment{}

\begin{document}


\title{A Deeply Buried Narrow-Line Seyfert 1 Nucleus Uncovered in Scattered Light}


\author{Xiang\ Pan\altaffilmark{1,2}, Honglin Lu\altaffilmark{2},
S. Komossa\altaffilmark{3}, Dawei Xu\altaffilmark{4}, Weimin Yuan\altaffilmark{4}, Luming Sun\altaffilmark{2}, Paul S. Smith\altaffilmark{6}, Shaohua Zhang\altaffilmark{1}, Peng Jiang\altaffilmark{1}, Chenwei Yang\altaffilmark{1}, Wenjuan
Liu\altaffilmark{7}, Ning Jiang\altaffilmark{3}, Y. E. Rashed\altaffilmark{8},
A. Eckart\altaffilmark{5}, Jens Dierkes\altaffilmark{5}, and Hongyan
Zhou\altaffilmark{1,2}} 
\altaffiltext{1}{Polar Research Institute of
China, 451 Jinqiao Road, Shanghai, 200136, China;
zhouhongyan@pric.org.cn}
\altaffiltext{2}{Key Laboratory for
Researches in Galaxies and Cosmology, Department of Astronomy,
University of Sciences and Technology of China, Chinese Academy of
Sciences, Hefei, Anhui, 230026, China}  
\altaffiltext{3}{Max Planck Institut f{\"u}r Radioastronomie, Auf dem H{\"u}gel 69,
D-53121 Bonn, Germany}
\altaffiltext{4}{National Astronomical Observatories, Chinese Academy of
Sciences, Beijing 100012, China}
\altaffiltext{5}{University and City Library Cologne,
Department Research and Publication Support, University of Cologne, Universit{\"a}tsstr, 33,
50931 Cologne, Germany}
\altaffiltext{6}{Steward Observatory, The University of Arizona, Tucson, AZ 85721, USA}
\altaffiltext{7}{Yunnan Observatories,
Chinese Academy of Sciences, Kunming, Yunnan 650011, China }
\altaffiltext{8}{Department of Astronomy, Faculty of Science, University of Baghdad, 10071 Baghdad -- Aljadirya, Iraq}

\begin{abstract}

We present spectropolarimetric and spectrophotometric
observations of the peculiar active galactic nucleus (AGN) SDSS
J120300.19+162443.7 (hereafter J1203+1624) at $z=0.1656$. Its
optical total flux spectra clearly show broad emission lines (BELs)
in H$\alpha$ and H$\beta$. After removal of narrow emission lines 
(NELs), the full width at half maximum (FWHM) of the Lorentzian BEL is 
$FWHM_{\rm BEL} \approx 1270$~\kms, fulfilling the conventional
definition of a narrow-line Seyfert 1 (NLS1) galaxy. However, its
NELs are unprecedentedly strong when compared to type 1
AGNs. This, together with its
large MIR excess ($g-W_4 = 13.172$ mag), implies that the
observer and the NEL region might see a
different ionization continuum. Our optical spectropolarimetry confirms its type 2 nature by detecting a polarized blue continuum and Balmer BELs ($FWHM_{\rm Polarized~BEL} \approx 1,183$~\kms),
 with a high polarization degree of $> 20$\% in the blue wing.
The mass and Eddington rate of the central black hole are estimated based on both transmitted and scattered AGN radiation, which is $M_{\bullet} < 2.9 \times 10^7 M_\odot$ and $L_{\rm bol}/L_{\rm Edd} > 1.5$. Severe extinction of the AGN emission also enables a clear view of the compact host galaxy. 
The discovery of J1203+1624 suggests that NLS1 follows the AGN unfication scheme, and studying of its analogs could
blaze a new trail for exploring the connection between black hole growth and star formation in the host galaxy. 
The interesting features of J1203+1624, like the peculiar NELs and inflowing scattering clouds within the sublimation radius, are worth detailed follow-ups in the future.
\end{abstract}
%and are even exceptional for type 2s. 
%Optical \feii emission line multiplets, which are often strong in NLS1s, are also detected in the total flux spectra.
%Together with its large MIR excess ($W_1-W_4 = 8.48$)



\section{Introduction}

In the 1990s, an orientation-based unified model was proposed that explains the spectroscopic classification of active galactic nuclei (AGNs; Antonucci 1993). In this toy model, an AGN consists of an accretion disk near the black hole, a fast-rotating broad emission-line region (BELR), and a more distant narrow emission-line region (NELR) in the ionization cone that produce the continuum, broad emission lines (BELs) and narrow emission lines (NELs), respectively. If the observers line of sight lies within the ionization cone, then a type 1 AGN is observed where the disk continuum and the emission from the BLR and NLR are all in direct view. A dusty torus is also assumed that, if viewed edge-on, will block the disk continuum and BELR in our line of sight, giving rise to ``type 2'' AGNs such as a Seyfert 2 galaxy (Peterson 1997).

The term narrow-line Seyfert 1 (NLS1) galaxy was initially denominated by
Osterbrock \& Pogge (1985) to categorize an ``amazing'' class of AGN (Tarchi
et al. 2011) that shows a prominent disk continuum, but their broad
emission-line widths are much narrower than those for normal Seyfert 1 nuclei
(BLS1s).
There are other observational differences between NLS1s and
BLS1s, including: (1) \feii emission lines in
NLS1s are often stronger, (2) the BEL profile of NLS1s is
often better described by a Lorentzian than a Gaussian function
(e.g., V{\'e}ron-Cetty et al. 2001; Zhou et al. 2006), (3) NLS1s
more frequently show a steep photon index and short-time-scale/large-amplitude 
variability in the X-rays (e.g., Leighly 1999a; 1999b),
and (4) the fraction of radio-loud NLS1s is much less than that of
BLS1s (Zhou et al. 2002; Komossa et al. 2006; Zhou et al. 2006; Yuan
et al. 2008). It is now generally thought that these extreme
properties of NLS1s result from smaller supermassive black
holes (SMBHs) accreting at higher Eddington rates than BLS1s.

On the other hand, it has been proposed that the narrower BELs in
NLS1s come from a disklike BELR viewed at a small inclination angle
from face-on (Osterbrock \& Pogge 1985; Leighly 1999b; Komossa et
al. 2006). This is supported by the fact that at smaller inclinations,
the observed emission-line width of quasars are relatively smaller(e.g. Wills \& Browne 1986; Rokaki et al. 2003).
Also, the absence of absorption by materials of the BELR in type 1 quasars (Antonucci et al. 1989) 
and the modeling of the velocity-resolved reverberation mapping results 
(e.g. Grier et al. 2013, Li et al. 2015) both
favor a disklike BELR geometry. However, this geometrical interpretation alone cannot explain 
all of the extreme properties of NLS1s (Reviewed in Komossa 2008). This is supported by 
the spectropolarimetric study of the orientation of NLS1s (Robinson et al. 2011).
%Shen \& Ho 2014
Nevertheless, both our current samples and interpretations of NLS1s prefer small inclination angles,
from which the continuum and BELs are too strong for analysis of weak structures like NELRs, host galaxies, etc.
If NLS1s follow the ``strong'' unification model of AGNs, there should be their type 2
counterparts (Narrow Line Seyfert 2s, or NLS2s), which are viewed 
nearly edge-on. Studying them may reveal hidden properties
that can be crucial in understanding AGNs with extreme properties, such as NLS1s.

There are studies devoted to identifying NLS1s along AGN unification
sequences, including NLS2s and intermediate-type counterparts of
NLS1s (NLS1.8s or 1.9s; Osterbrock 1981). Zhang et al. (2017b)
recently presented a detailed analysis of three optically classified
NLS1s with \ebv$\sim$ 1 mag reddened by dusty absorbers at torus
scales.
As extinction grows stronger, BELs can only be detected in
the IR because of relatively less extinction at longer wavelengths.
The \pabeta~BELs of $FWHM<2000$ \kms are reported in two optically
obscured Seyfert galaxies, NGC 5506 (Nagar et al. 2002) and Mrk 573
(Ramos Almeida et al. 2008), making them good candidates for moderately
buried NLS1s. However, in type 2 AGNs, where dust extinction is much
stronger, transmitted BELs are so weak that measurements of line
width and verification of their NLS1 nature becomes difficult.
Dewangan \& Griffiths (2005) reported three type 1.9/2 Seyfert
galaxies with NLS1-like X-ray properties and suggested them as
NLS2s. Based on either IR BELs or X-ray properties, these candidates
are identified differently from the conventional NLS1s, and their
hidden BELs cannot be easily probed. 
%\x{During the establishing of the unification model, } 
In this paper, we report the
discovery of a bona fide NLS2, SDSS J120300.19+162443.7 (hereafter
J1203+1624). We identified it as a NLS1 based on its total flux
optical spectrum, while its type 2 nature is implied by its superstrong
NELs and extremely high flux excess in the FIR. Our follow-up optical 
spectropolarimetry revealed the scattered BEL component and confirmed its NLS2 nature.

\section{Observations}
With an archival SDSS (York et al. 2000) spectrum (3100--9180\AA)
observed on 2007 Apr 17, J1203+1624 is identified as a NLS1 in our
systematic study of SDSS quasars (\S3). This spectrum is combined with the UV spectrum obtained on 2006 March 28 by $GALEX$ (Morrissey et al. 2007), and broad-band
photometry from $GALEX$ (UV), SDSS (optical), 2MASS (near-IR; the Two
Micron All Sky Survey, Skrutskie et al. 2006) and $WISE$ (mid-IR; Wide-field
Infrared Survey Explorer, Wright et  al. 2010) is collected
to study the spectral energy distribution (SED) of J1203+1624.

The strong NELs are very unusual for a NLS1.
We followed up by observing it using the 6.5m MMT with the blue channel spectrograph
(Schmidt et al. 1989) on 2011 Dec 27 (G800 grating, [3350--5100]\AA),
and in 2012 Feb 29 (G600 grating, [8770--9520]\AA). We also performed
optical spectroscopy of J1203+1624 with the MODS spectrograph on
the Large Binocular Telescope (LBT\footnote{The LBT is an
international collaboration among institutions in the United States,
Italy, and Germany. The LBT Corporation partners are the University of
Arizona, on behalf of the Arizona Board of Regents; Istituto
Nazionale di Astrofisica, Italy; LBT Beteiligungsgesellschaft,
Germany, representing the Max-Planck Society; the Leibniz Institute
for Astrophysics Potsdam, and Heidelberg University; Ohio State
University, and the Research Corporation, on behalf of the
University of Notre Dame, University of Minnesota, and University of
Virginia.}; 3200--10200\AA) on 2012 Feb 01 (Rashed et al. 2015). 
The SDSS, MMT/BC, and LBT/MODS spectra
are in similar resolutions. After scaling the three spectra at
a similar continuum flux level,
no obvious variability in their spectral shape
is found. This is consistent with the stable $V$-band
magnitude of 18.10$\pm$0.11 monitored by the Catalina Sky Survey
\footnote{http://nesssi.cacr.caltech.edu/DataRelease/} between
2005 Apr 9 to 2013 Jun 11 (Catalina ID: CSS\_J120300.2+162444). In
addition, a medium-resolution NIR spectrum (9200-25000\AA) was
obtained on 2013 Feb 23 with the TripleSpec spectrograph on the
Palomar 200-inch Hale telescope, which helps in diagnose its
emission-line systems. And optical spectropolarimetry was carried out
with MMT/SPOL (Schmidt et al. 1992) on 2016 Apr 5 to search for
scattered emission in the optical band. With a 600 g mm$^{-1}$ grating, 
the SPOL data cover a wavelength range of [4200,8200]\AA.
 Since variability is not
detected, and the optical-IR spectra of J1203+1624 have
similar resolutions, they are combined to reach a high signal-to-noise 
ratio (SNR) and
a wide wavelength coverage. All data are corrected for the Galactic
extinction of \ebv=0.034 according to the dust map in Schlafly \&
Finkbeiner et al. (2011), and deredshifted with $z = 0.16559$
determined by the NELs before follow-up analysis. Throughout this
paper, we adopt a cosmological model with parameters $H_0=70$~\kms
Mpc$^{-1}$, $\Omega_{\rm M}=0.3$, and $\Omega_{\Lambda}=0.7$.

%\section{Optical Spectral modelling}
\section{An Unusual NLS1}

Following Zhou et al (2006), we carry out a two-step iterative
approach  to  fit both the continuum and emission lines of the 
optical spectrum of J1203+1624. 
Here is a brief summary. In step 1, the  optical continuum  is  fitted  in
continuum windows with a combination of host  starlight and nuclear
emission. The simple stellar population (SSP) models in Bruzual \&
Charlot (2003) are  assumed as host starlight, reddened  with a
Milky Way (MW) extinction law (Fitzpatrick \& Massa 2007). The
nuclear emission includes a power-law (PL) continuum and a  two-component
analytic \feii emission model (Dong et al. 2005) that incorporates the
VJV04 \feii template (V{\'e}ron-Cetty et al. 2004). During continuum fitting, 
spectral regions around strong stellar absorption features (gray regions in 
the lower panel of Figure~1) are weighted higher than the rest of the data points 
so as to ensure a reasonable fit to the SSP component. The fitted
continuum model is then subtracted, and the line spectrum is
obtained. In step 2, the line spectrum derived in the first step is fitted.
Besides the prominent NELs, a broad base is clearly seen at \halpha, 
indicating the presence of a (low-ionization) BEL component. A test fitting of the \halpha+\nii+\sii
complex shows that the NELs are in the form of a single Gaussian, while a Lorentzian BEL can
fit the data better than a Gaussian BEL. 
%The NELs are in the form of gaussians, while a Lorentzian profile fit the BEL system better than Gaussian profile ($\Delta\chi^2/{\rm DOF} = \frac{\chi^2_{\rm Gaussian Model} - \chi^2_{\rm Lorentzian Model}}{\rm Degree of Freedom} = 1.86$).
Thus, each of the identified NELs/BELs in the line spectrum is assigned a Gaussian/Lorentzian profile, with their line centroids and widths bound together in velocity space. Then the line spectrum is modeled. In step 3, we repeat the fitting of the continuum and emission lines until the model parameters converge.
The converged continuum and emission-line models are displayed in Figure~1.
Finally, we resampled the spectral data and carried out the fitting for 1,000 times with 
a bootstrap approach. The results are then compared in obtaining uncertainties of model parameters. The modeled fluxes of each NEL/BEL are listed in Table~1.

The spectral modeling reveals a blue nucleus, with a power-law at
$\alpha_\lambda = -1.36\pm0.24$. The widths of the Lorentzian BELs are $FWHM_{\rm BEL} = 1270\pm18~$ \kms, four times the width of the gaussian NELs of $FWHM_{\rm NEL} = 305\pm1$\kms, making the narrow- and broad-line components relatively easy to separate. Thus, J1203+1624 fulfills the criterion of $FWHM_{\rm BEL} < 2200~$ \kms for NLS1s (Zhou et al. 2006; Gelbord et al. 2009). A
Balmer flux ratio of the Lorentzian BEL $F_{\rm \halpha BEL}/F_{\rm \hbeta BEL} = 2.79\pm$0.43 is found,
consistent with the sample mean of 3.028$\pm$0.017 for NLS1s (Zhou et al.
2006). Another typical component for NLS1s, \feii emission, is
marginally detected in J1203+1624, with a relative intensity $R_{4570}
\equiv F_{\rm \feii \lambda\lambda4344-4684}/F_{\rm \hbeta} = 
0.25\pm0.05$. The total flux ratio \oiii $\lambda$ 5007/\hbeta %$_{\rm NEL+BEL}$
is found to be 4.45$\pm$0.26, which exceeds the upper
limit of 3 adopted for NLS1s. 
In early works, the \oiii $\lambda$ 5007/\hbeta $<$ 3 criterion is to ensure 
the existence of a Balmer BEL system in the nuclei of NLS1 candidates. 
The BEL system in J1203+1624 is directly observed, 
and the detection of \feii multiplets and high-ionization \fevii, \fex emission lines suggests the 
presence of the AGN. Therefore, this line ratio diagnosis is not prerequisite (Pogge et al. 2000,
Zhou et al. 2006), and J1203+1624 is identified as a NLS1 by any standard.

As shown in Figure~2, the Balmer decrements of NELs in J1203+1624 are found to be in an excellent agreement with that of the Case-B predication (Osterbrock \& Ferland 2006), which means negligible dust extinction to the NELR in J1203+1624. We then checked J1203+1624 in the BPT diagrams (Figure~3) for an empirical diagnose of the narrow emission lines. 
Given the presence of various strong high-ionization lines, the NEL of J1203+1624 is expected to be AGN-dominated.  However, J1203+1624 is unexpectedly diagnosed as a composite of AGNs and star-forming (\hii) regions.
With a similar \oiii/\halpha ratio to AGNs, the ionization source of the NELR can be AGN-dominated in J1203+1624. On the other hand, the relatively low  \nii/\halpha ratio may indicate a relatively low metallicity of the NELR.
The most unusual property of the NELs in J1203+1624 as a NLS1, though, is the high intensity.
At $EW_{\rm \oiii \lambda 5007} = 583\pm16$\AA, it is even higher than 97\% of SDSS DR1 type 2 quasars (Zakamska et al. 2003). The qustion is raised as to whether or not J1203+1624 is a Seyfert 2 galaxy. 
We thus check the spectral energy distribution of J1203+1624 in a broader wavelength range for additional clues in \S4.

\begin{figure*}
\includegraphics[width=1.0\textwidth]{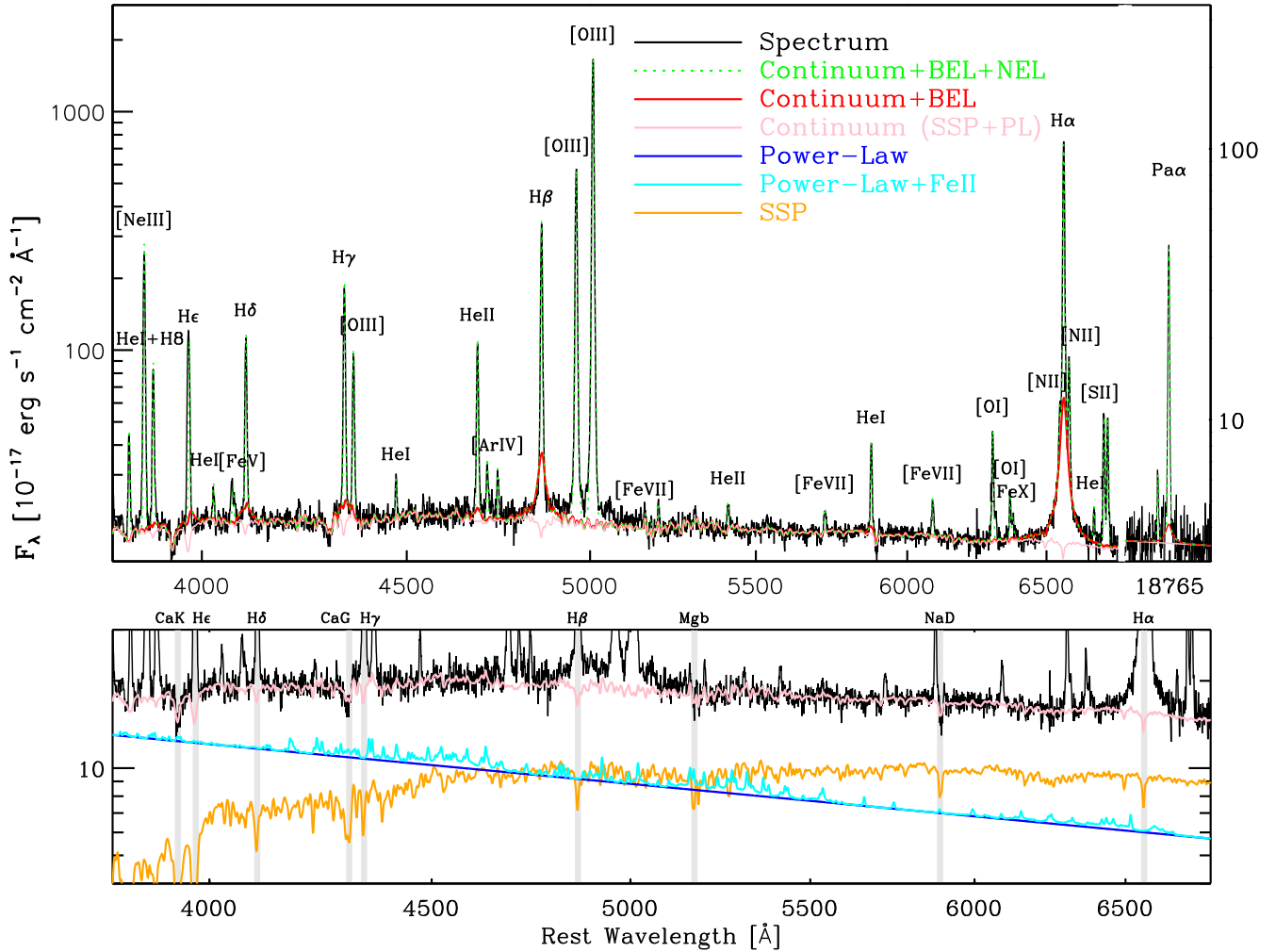}
\caption{Spectral modeling of J1203+1624. The observed data are
shown with black lines. The best-fit models, i.e. continuum, continuum+BEL, and continuum+BEL+NEL, are
overlaid in different colors in the upper panel. In the lower panel, 
a zoom-in comparison between the data and modeled continuum (PL+\feii+SSP) components is shown,
with prominent stellar absorption lines marked with vertical gray lines. }
\end{figure*}

\begin{figure}
\includegraphics[width=0.48\textwidth]{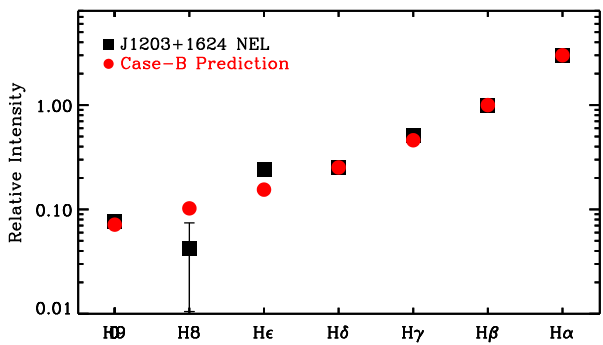}
\caption{Excellent agreement is found between the
Balmer decrements of NELs in J1203+1624 and the Case-B model prediction, which
indicates negligible dust extinction to the NELR.}
\end{figure}

\begin{figure}
\includegraphics[width=0.48\textwidth]{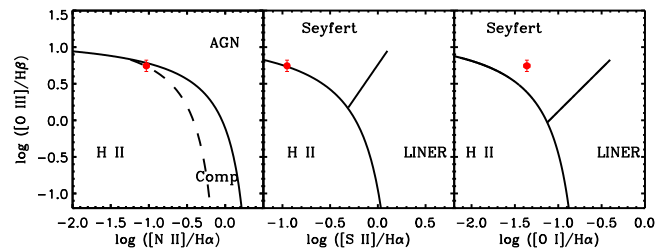}
\caption{The NEL ratios of J1203+1624 in the BPT diagrams (Baldwin et al. 1981; Kewley et al. 2001; Kauffmann et al. 2003), which is diagnosed as a composite of AGN and star-forming (\hii) regions.}
\end{figure}

\begin{deluxetable}{lccc}
\tablenum{1}
\tablecaption{Emission Line Intensities}
\tablewidth{0pt}
\tablehead{
\colhead{ID} & \colhead{$\lambda_{lab}^{\rm (a)}$} & \colhead{$L$/10$^{40}$ erg s$^{-1}$}
}
%\colhead{ID} & \colhead{$\lambda_{lab}$} & \colhead{L 10$^40$ erg s$^{-1}$}}
\startdata
 NELs$^{\rm (b)}$ & & \\ \hline
H9           &     3836.47&   6.8$\pm$   0.3\\
\neiii       &     3869.85&  60.9$\pm$   1.0\\
\hei         &     3889.74&  13.1$\pm$   4.2\\
H8           &     3890.15&   3.8$\pm$   4.0\\
\hepsilon    &     3971.20&  21.5$\pm$   2.8\\
\hei+NII     &     4027.28&   2.0$\pm$   0.4\\
\fev         &     4072.39&   2.1$\pm$   0.1\\
\hdelta      &     4102.89&  22.6$\pm$   2.8\\
\hgamma      &     4341.68&  45.1$\pm$   3.1\\
\oiii        &     4364.44&  21.3$\pm$   0.4\\
\hei         &     4472.76&   2.8$\pm$   0.7\\
\heii        &     4687.02&  25.2$\pm$   0.6\\
\hei         &     4713.15&   4.6$\pm$   0.6\\
\ariv        &     4741.50&   3.6$\pm$   0.4\\
\hbeta       &     4862.68&  88.6$\pm$   3.5\\
\oiii        &     4960.29& 162.5$\pm$  16.5\\
\oiii        &     5008.24& 492.2$\pm$  50.1\\
\fevii       &     5160.33&   0.5$\pm$   0.3\\
\fevi        &     5177.48&   0.5$\pm$   0.1\\
{[N\,{\footnotesize I}]}&     5200.53&   1.4$\pm$   0.2\\
\fevii       &     5277.85&   0.0$\pm$   0.1\\
\fexiv       &     5304.34&   0.4$\pm$   0.1\\
\heii        &     5413.60&   1.7$\pm$   0.4\\
\fevii       &     5722.30&   1.6$\pm$   0.2\\
\hei         &     5877.29&   8.4$\pm$   0.4\\
\fevii       &     6087.98&   2.7$\pm$   0.2\\
\oi          &     6302.05&  11.6$\pm$   0.5\\
\oi          &     6365.54&   3.6$\pm$   0.2\\
\fex         &     6376.30&   1.4$\pm$   0.2\\
\nii         &     6549.85&   8.2$\pm$   0.4\\
\halpha      &     6564.61& 266.8$\pm$   6.1\\
\nii         &     6585.28&  24.7$\pm$   1.1\\
\hei         &     6679.70&   2.4$\pm$   0.2\\
\sii         &     6718.29&  15.3$\pm$   0.4\\
\sii         &     6732.67&  14.5$\pm$   0.4\\ \hline
BELs$^{\rm (c)}$ & & \\ \hline
H8           &     3890.15&   0.5$\pm$   1.2\\
\hepsilon    &     3971.20&   3.8$\pm$   2.6\\
\hdelta      &     4102.89&   7.5$\pm$   3.3\\
\hgamma      &     4341.68&   4.6$\pm$   5.8\\
\heii        &     4687.02&   4.1$\pm$   1.9\\
\hbeta       &     4862.68&  40.1$\pm$   3.4\\
\hei         &     5877.29&   1.8$\pm$   0.8\\
\halpha      &     6564.61& 112.6$\pm$   9.7 
\enddata
\tablecomments{\\
(a) References for the identified emission lines are Liu et al.(2000), Vanden Berk et al. (2001);\\
(b) NELs (narrow emission lines) are in the profile of a gaussian, with a width of $FWHM_{\rm NEL} = 305\pm1$\kms;\\
(c) BELS (barrow emission lines) are in the profile of a gaussian, with a width of $FWHM_{\rm BEL} = 1270\pm18$\kms. }
\end{deluxetable}

\section{Obscured and Scattered Nuclear Emission}
\subsection{The Deeply Buried Nucleus}

As shown in Figure~4, the $GALEX$ data in the UV ($\lesssim 2500$ \AA~ 
in the quasar's rest frame) appear to be the
natural extent of the nuclear emission model described in \S3,
while the MIR-to-optical color of J1203+1624 is $g-W_4=13.172\pm0.050$ mag, 
2.3 mag higher than the quasar composite, and indicates a significant 
flux excess in the MIR. A similar MIR flux excess is found in the quasar SDSS
J000610.67+121501.2 (SDSS J0006+1215), which turns out to be heavily obscured AGN radiation (Zhang et al. 2017a).

We then modeled the UV-optical-IR spectral energy distribution (SED)
of J1203+1624 with three components, including (1) a blue nuclear
emission ($B_\lambda$) that dominates in the UV, (2) a reddened AGN
emission ($R_\lambda$) as indicated by the MIR flux excess, and (3)
starlight from the host galaxy ($H_\lambda$) which contributes
significantly in the optical and NIR (\S3). Since the spectral model of
the blue nucleus obtained in \S3 is similar to the quasar composite
in both their spectral indices and Balmer decrements, we follow Zhou et al. (2010) in applying the broadband
composite quasar spectrum ($Q_\lambda$) as blue nuclear emission $B_\lambda$ = $a$ $Q_\lambda$,
which largely simplifies our SED models. We assume $_\lambda$ = $b\
10^{-0.4 A_\lambda^{\rm nucleus}}$ $Q_\lambda$ as the reddened AGN
emission. Since a steep extinction law is favored for AGNs from recent
observations (e.g., Jiang et al. 2013; Zafar et al. 2015), the extinction law for the Small
Magellanic Cloud (SMC; Martin et al. 1989) is adopted as $A_\lambda^{\rm nucleus}$. 
Similar to \S3, for the stellar emission $H_\lambda$, reddened SSP spectra are applied. 
Finally, the photometric and spectroscopic total fluxes of J1203+1624 are fitted with this SED model 
($F_\lambda = \rm R_\lambda  + H_\lambda + B_\lambda$).

\begin{figure*}
\includegraphics[width=1.0\textwidth]{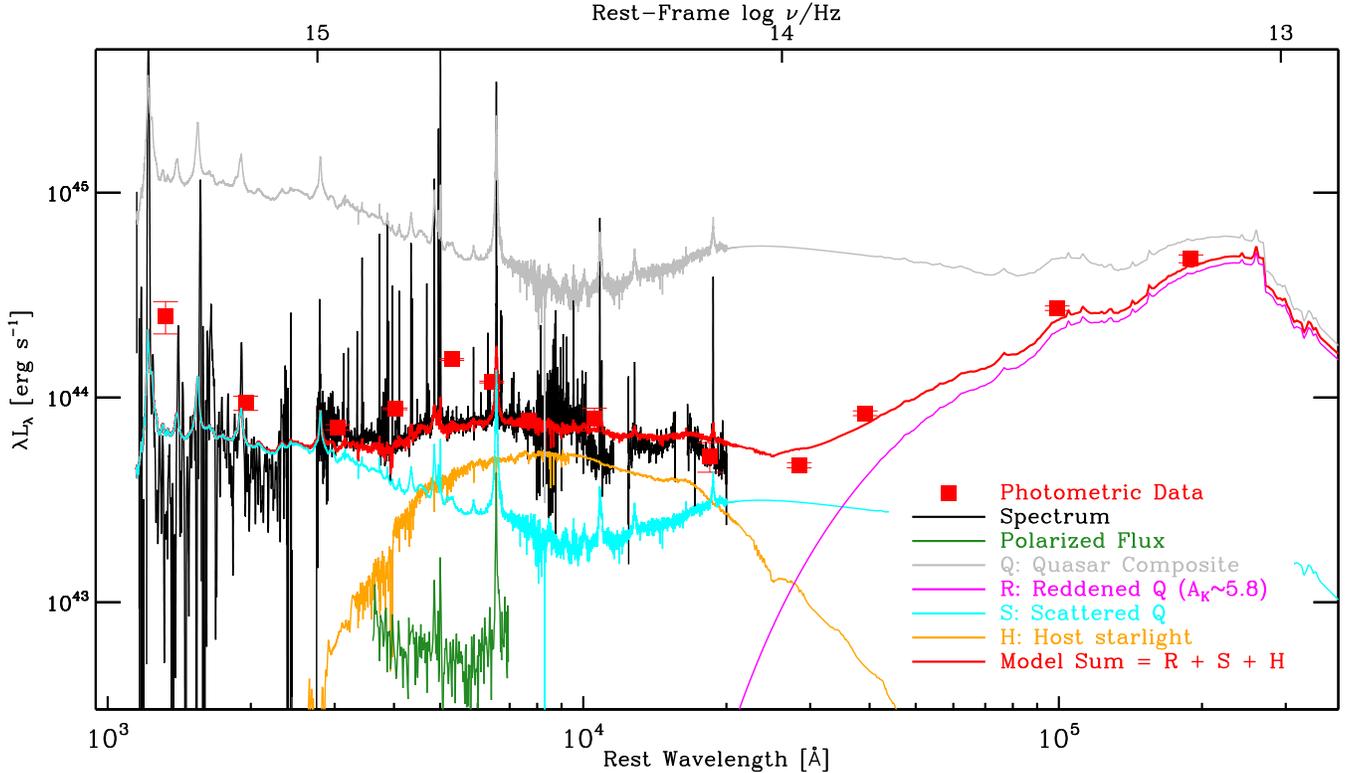}
\caption{The SED modeling of J1203+1624. The black line is the
combined spectrum, and red squares are the photometric data. They are
fitted with three components: a reddened quasar composite (magenta
line), which dominates in the MIR; the starlight of the host galaxy (orange line),
which contributes significantly in the IR-optical; and a scattered
quasar composite (cyan line), which dominates in the UV-optical. The
best-fit model is plotted with a red line, and the polarized flux is
presented with a green line.}
\end{figure*}

Fitting results suggest that the AGN emission is deeply buried in
dust with a color excess $A_K$ = 5.83$\pm$0.25 if the SMC extinction
law is assumed. This is consistent with a significantly red $WISE$ color of
$W_1 - W_4 = 8.48$, 2.42 mag higher than that of the quasar composite. 
At this level of high extinction, the transmitted
spectrum is negligible even in the NIR, which suggests that 
J1203+1624 is a Seyfert 2 galaxy. We checked the
line profile of the hydrogen line with the largest wavelength in our
spectrum, i.e. \paalpha, which is dominated by the NEL component
(right end of upper panel of Figure~1), consistent with the SED modeling
 result of no obvious transmitted BEL emission in the NIR. Since that the NELR is almost free from reddening (Figure~2), 
the obscurer should be located between the NELR and BELR and is very likely 
the dusty torus presumed by the AGN unification models. 
After extinction correction, the luminosity of this buried nucleus 
is found to be $L_{\rm bol} =$ 5.7$\pm1.2 \times10^{45} \rm erg\ s^{-1}$. 
This estimated bolometric luminosity is $\sim$1000 times that
of $L_{\rm \oiii\lambda5007}$ (Table~1), similar to the sample mean ratio of 
3400 (1$\sigma$ range of $\sim$1000--10000) for the 
58 bright quasars at low-to-intermediate redshift (Pennell et al. 2017).
The blue nuclear emission in the UV is only $\sim~3.6 \% \pm 0.3\%$ of the reddened AGN radiation.
Such a weak blue nuclear emission is also discovered in the high-polarization reddened quasar OI 287
(Goodrich \& Miller 1988; Li et al. 2015) and the heavily obscured quasar SDSS J0006+1215
(Zhang et al. 2017a), where it is interpreted as scattered AGN
radiation. To find out whether reflection also plays a
role in J1203+1624, we performed spectropolarimetric
observations with MMT/SPOL.

\subsection{Scattered AGN Radiation}

In previous works, spectropolarimetry plays an important role in 
revealing reflected nuclear emission in type 2 AGNs, such as NGC 1068 
(Antonucci \& Miller 1985), Mrk 477, Mrk 1210, NGC 7212, and Was 49b (Tran et al. 1992).
 These observations are among the key motivations of the AGN unification model. 
Enlightened by these works, it is realized that the buried NLS2 nuclear
can possibly be probed and identified via spectropolarimetry of scattered light. 
Previous spectropolarimetric studies of NLS1s do find significant scattered nuclear 
radiation in several NLS1s, like Mrk 766, Mrk 1239, etc (detailed sample studies 
can be found in Goodrich et al. 1989, Robinson et al. 2011), which suggests that 
scattering can be common in high-accreting AGNs like NLS1s.

\begin{figure*}
\includegraphics[width=\textwidth]{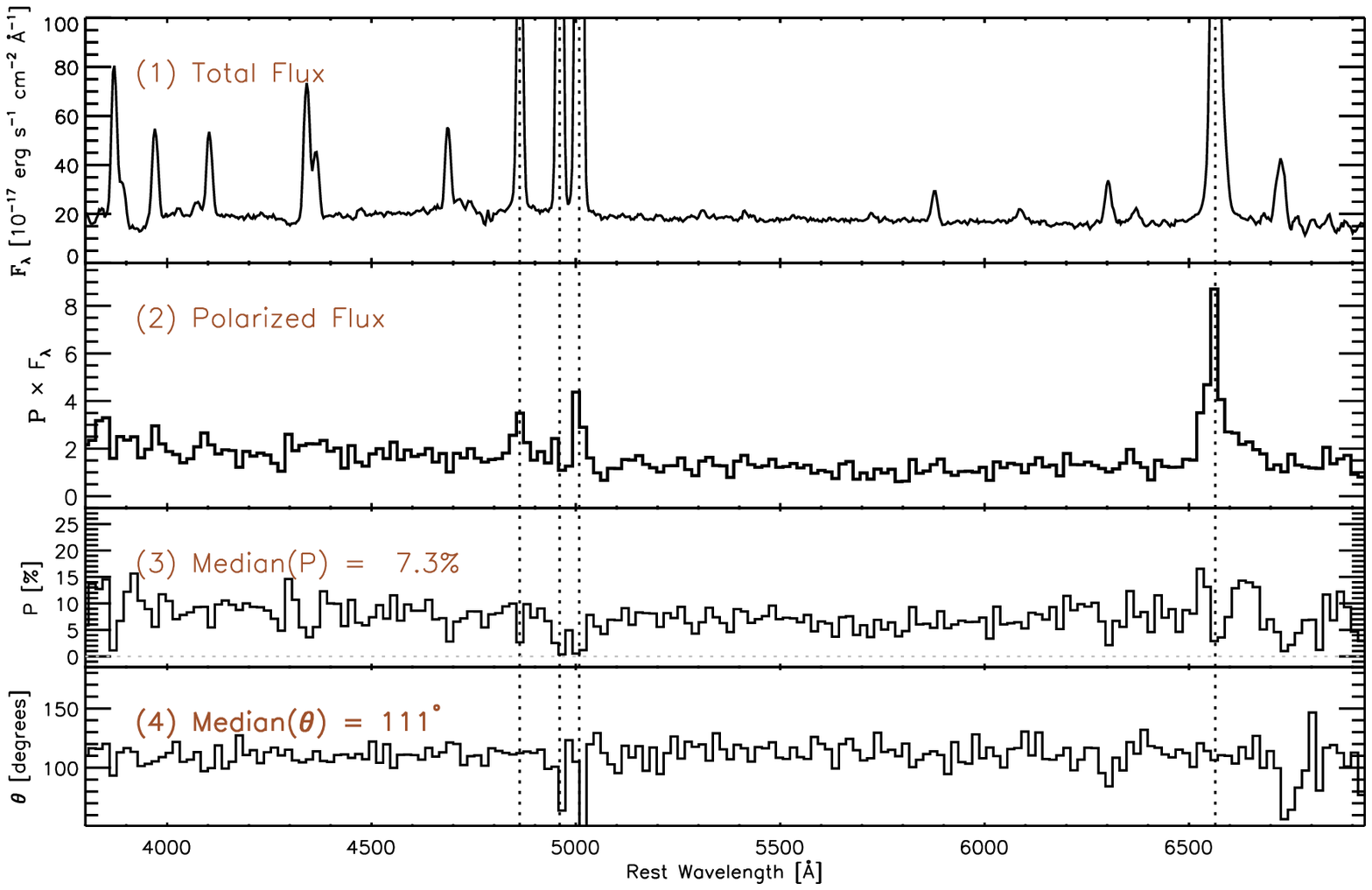}
\caption{The total and polarized flux spectra ($P\times F_\lambda$) of
J1203+1624 are shown in panel (1) and panel (2) respectively.
In panel (3), degree of polarization ($P$) is shown as a function
of wavelength, with a median value of 7.3\%. The polarization angle ($\theta$) is displayed in the panel (4), which is around 111$^\circ$ for both continuum and BELs. The data in this figure is binned every 5 pixels.}
\end{figure*}

\begin{figure}
\includegraphics[width=0.48\textwidth]{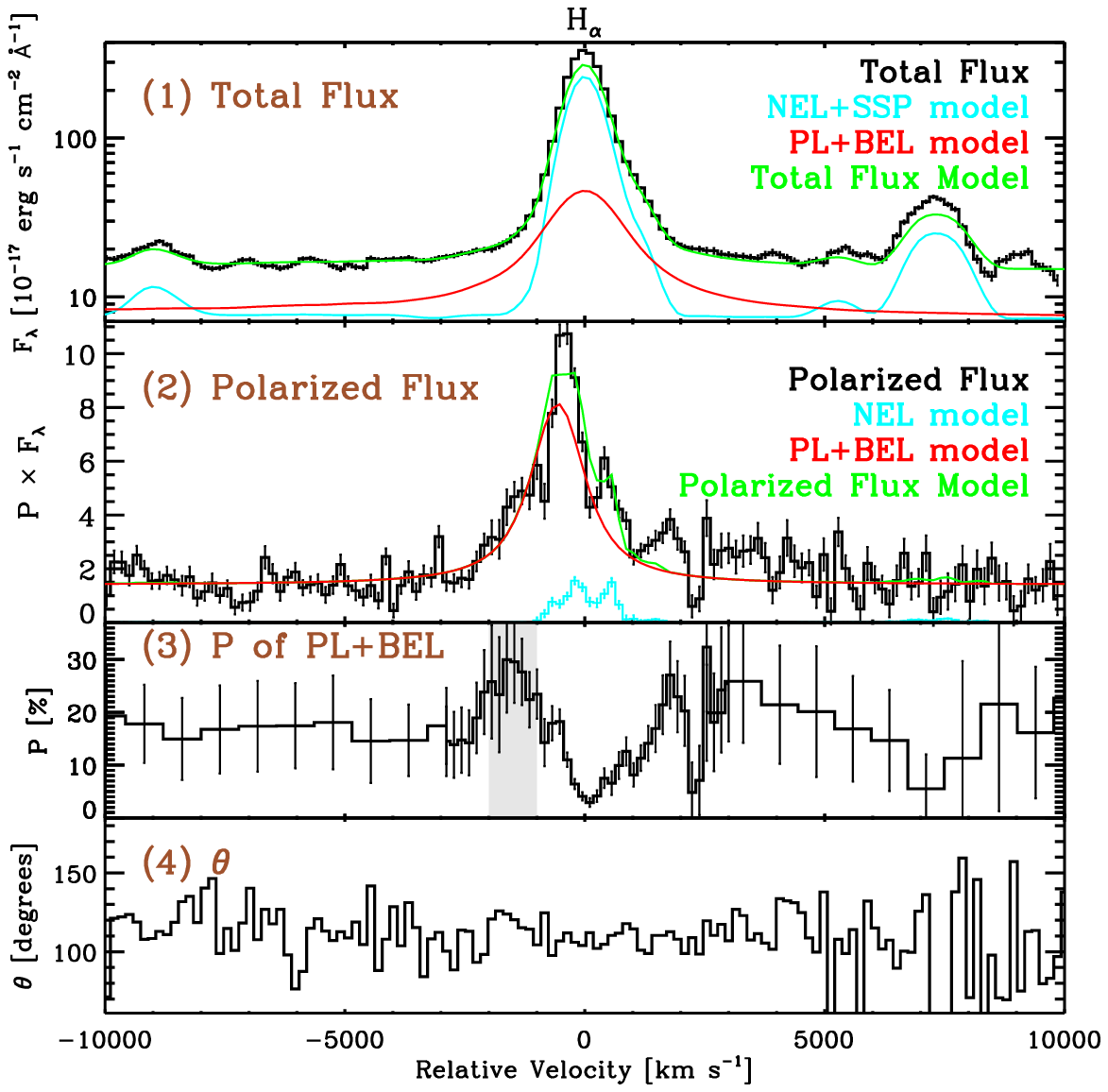}
\caption{The total and polarized flux spectra ($P\times F_\lambda$) of
J1203+1624 around \halpha region in velocity space are shown in panel (1) and panel (2) respectively.
The models of NEL(+SSP), power-law+BEL and their sum are shown in cyan, red and green 
respectively for both the total flux and polarized flux spectrum. After the decomposition,
degree of polarization $P$ for the power-law+BEL component is obtained in panel (3), notice that the data at $|v| > 3000~$\kms\ are binned every 5 pixels.
Notice enhanced polarization in the blue wing ($v \in [-2000,-1000]~$\kms~, 
grey-shaded) of \halpha. The polarization angle is displayed in panel (4).
}
\end{figure}

A degree of polarization of $P \sim 7.3$\% (Figure~5(c)) is found in the SPOL data of J1203+1624, higher than 90\% of type II AGNs reported in Miller \& Goodrich (1990), Zakamska et al. (2005), and Ramos Almeida et al. (2016),
indicating that a significant portion of the optical light comes from scattering. 
Besides, the H$\alpha$ and H$\beta$ BELs are significantly detected in the 
polarized spectrum ($P\times F_\lambda$; Figure~5(b)). And, a universal polarized 
angle of $\theta \sim 111^\circ$ for both continuum and BEL (Figure 5(d)) suggests a similar 
origin for them (similar to 3C234, reported in Antonucci 1984), while the NELs are having lower $P$ and a 
different polarized angle, indicating a different origin of polarization. 
Since the polarized radiation from the core components (PL continuum and BEL) comes of a similar origin, we follow Antonucci 1984 in decomposing them from both the total flux and the polarized flux spectrum to analyze the property of its polarization.
After been convolved at a resolution of $\sim 7~$\AA, the spectral model (green dotted line in Figure~1; see \S3 for a detailed description) roughly reproduces the total flux spectrum observed by MMT/SPOL (Figure~6(a)). Similarly, the convolved AGN PL+BEL model (red solid line) and NEL+SSP model (cyan solid line) are also obtained. It is assumed that there is no polarization in starlight, since it is much more extended as compared to the nucleus or AGN polar scatterers. This leaves us with three components to consider in the polarized flux, AGN power-law continuum, BELs and NELs. Since the strongest isolated NEL in the polarized flux spectrum is \oiii $\lambda$ 5007, we apply its profile in generating the polarized NEL model. On the other hand, as dust extinction seems to be negligible in the NELR (Figure~2) and no apparent wavelength dependence is seen in $P$, electron scattering is therefore assumed here. The NEL flux ratios derived in the total flux spectrum (Table~1) are also applied in obtaining the polarized NEL model (cyan line in Figure 6(b)). After subtracting the polarized NEL model in the polarized flux spectrum, the residual is fitted with a PL and a Lorentzian BEL (red solid line in Figure~6(b)) in the \halpha region. The width of the polarized BEL is found to be $FWHM_{\rm Polarized~BEL} = 1183 \pm 77$~\kms, and the line core is blueshifted at $563 \pm 59~$\kms. In fact, the emitting regions of \halpha and the \oiii $\lambda$ 5007 NELs can be quite different (Baskin \& Laor 2005), and the assumption that the $P$ and the profile of polarized \halpha and \oiii $\lambda$ 5007 NEL are the same can be invalid. Nevertheless, this assumption provides us with chances to reasonably estimate the polarized NELs. After subtracting SSP and NELs, the PL+BEL components in both total flux spectrum and polarized flux spectrum are obtained. The degree of polarization, $P$, of the PL+BEL component is then obtained (Figure 6(c)).

% median P 13.6
% P in the blue wing [-2000,1000]kms, max 29.96, median=27.67
 In type 2 AGNs like J1203+1624, polar scattering is the simplest and best explanation of the polarized nuclear emission.  Similarly, the AGN PL continuum and BELs have relatively high $P$ and close polarization angles, they are probably scattered by clouds that posit in similar polar directions relative to the AGN. Since there does not seem to be a dramatic rise at the blue end of the polarized flux spectrum in J1203+1624, electron scattering is therefore favored\footnote{The dust scattering cross-section rises sharply at short wavelengths (Mathis et al. 1977).}.
As mentioned earlier, the SED modeling results in a flux ratio of $\sim$3.6\% between the blue AGN and the buried AGN components, which indicates a similar value for the covering fraction of the scattering clouds if the bulk of the blue AGN flux comes from scattering. By assuming the incident line emission of the scattering cloud to be monochromatic, i.e. the polarized BEL is entirely broadened by the thermal motion of electrons, the upper limit of electron temperature is then $T_{\rm max} = \frac{m_e FWHM_{\rm polarized\ BEL}^2}{16k\rm ln2} = 8330\rm K$ (Miller et al. 1991). The temperature of ions/atoms/molecules is even lower. In order to keep electron scattering over dust scattering at such a low temperature, the scattering cloud should be metal-poor as compared to normal ISM to make it dust-deficient.
At such a considerable covering fraction and a relatively close distance to the AGN, the scattering clouds should themselves produce emission lines. Since both the scattering clouds and the strong NELR of J1203+1624 reside in the ionization cone and appear to be metal-poor \& dust-deficient, they are plausibly associated with each other. On the other hand, the $P$ of AGN PL+BEL is higher in the blue wing of the BEL as shown in Figure 6(c). In the velocity range of $v \in [-2000,-1000]~$\kms~ around \halpha (gray shaded in Figure 6.(c)), the median($P$) of PL+BEL is 27.7\%, twice the median value for the continuum. Such enhanced $P$ values in the blue wing of the \halpha BEL suggests that there should be additional scattering clouds relatively close to the BELR (Smith et al. 2000, 2003).  The redshifted polarized BEL can be caused by the inflowing motion of the scattering clouds toward the black hole at the velocity $w = u/(1-cos<\phi>_p) \gtrsim 500 \rm km s^{-1}$, where $u = 563 \pm 59 \rm km s^{-1}$ is the shifting speed of the polarized emission line (see \S4.2) and $<\phi>_p$ is the average scattering angle. Being even closer to the illuminant, this additional part of the scatterer can be highly ionized inflows inside the sublimation radius, which could give rise to the prominent coronal NELs (Rose et al. 2015) in J1203+1624. 

\subsection{The Black Hole Parameters}

As a brief summary, the analysis of the SED and polarized flux spectrum strongly suggests that J1203+1624 is a type 2 AGN whose NLS1 nuclear emission is revealed in scattered light.
Based on SED modeling, the bolometric luminosity of J1203+1624 is $L_{\rm bol} = 5.7\pm1.2 \times10^{45} \rm erg s^{-1}$, and the flux ratio between the blue AGN and the buried nucleus is $\sim 3.6\% \pm 0.3\% $ (\S3).
Applying this flux ratio to the observed \hbeta BEL, the intrinsic \hbeta BEL of J1203+1624 can be over twice the flux of \oiii $\lambda$ 5007, quite normal now among NLS1s. The BEL line widths measured in the total flux spectrum ($FWHM_{\rm BEL} = 1270\pm18~$) and the polarized flux spectrum ($FWHM_{\rm Polarized~BEL} = 1183 \pm 77$~\kms) are similar to each other. However, these observed BELs are originiated either from scattering or leaked BEL radiation viewed near the equatorial plane. The velocity width measurement in J1203+1624 is thus different from the measurement in type 1 AGNs, since in either case, the thermal motion of the scattering electrons 
 or the rotational motion of the disklike BELR can both broaden the observed line width. Therefore, the BEL width we measured can only be treated as an upper limit. Based on the line width and corrected luminosity of the \halpha BEL, we apply the empirical relation of Greene \& Ho (2005) and obtain the upper limit of the black hole mass of J1203+1624, $M_{\bullet} <  2.9\times 10^7 M_\odot$. A lower limit of a super-Eddington rate is obtained, $L_{\rm bol}/L_{\rm Edd} > 1.5$.

\section{Host Galaxy}
As a NLS2, the active nucleus in J1203+1624 is deeply buried in dust, and the relatively weak starlight from the host galaxy is clearly viewed.
A reasonable fit of the broadband data in \S4 is obtained with a single SSP at an age of 1.4Gyr. At a stellar mass of $M_* \sim 7.7 \times 10^{9} M_\odot$, the luminosity of starlight $L_* \sim 1.4 \times 10^{10} L_\odot$ is only $\sim 1$ \% of the bolometric luminosity of the dereddened NLS1 nucleus. The SDSS image of J1203+1624 is PSF-dominated, i.e. the physical extent of the host galaxy is small considering the relatively low redshift of $z=0.1656$. The stellar velocity dispersion is found to be $\sigma_* = 99\pm31$ \kms. Following the parameterized $M_{\bullet}--\sigma_*$ 
relation (Tremaine et al. 2002), the predicted black
hole mass according to $\sigma_* = 99\pm31$ \kms\ is $M_{\bullet}' = 0.80^{+1.57}_{-0.58}\times10^7 M_\odot$, consistent with the upper limit estimated using the \halpha BEL.

\section{Conclusion and Prospect}
In this paper,  we report the discovery of an unusual NLS1, J1203+1624, with extremely strong NELs. The $EW_{\rm \oiii \lambda 5007} = 583\pm16~$\AA\ is among the top values, even among type 2 quasars. After detailed analysis of its SED, J1203+1624 is found to be a type 2 AGN. The nuclear emission is revealed through scattered or possibly leaked radiation in the optical band. The scattered or possibly leaked NLS1 nucleus is identified in the total flux spectrum ($FWHM_{\rm BEL} \sim 1270$\kms) with a \feii relative strength of $R_{4570} \sim 0.25$, and confirmed in the polarized flux spectrum ($FWHM_{\rm Polarized~BEL} \sim 1183$\kms) with a high polarization degree ($P \sim 7.3\%$). And the obscurer is at a scale similar to the dusty torus proposed in the unification model. Taking advantage of its heavy extinction and prominent scattering of the core, both the AGN and the host galaxy can be well explored. The black hole mass is estimated to be $M_{\bullet} < 2.9 \times 10^7 M_\odot$, with a super-Eddingtion ratio of $L_{\rm bol}/L_{\rm Edd} > 1.5$. A mid-aged SSP at an age of $\sim 1.4$Gyr is found, with a stellar mass of $M_* \sim 7.7 \times 10^{9} M_\odot$ and a stellar velocity dispersion of $\sigma_* = 99\pm31$ \kms. A comparison of the nucleus and host galaxy suggests that J1203+1624 largely follows the $M_{\bullet}-\sigma_*$ correlation. Other interesting facts about J1203+1624, like the peculiar NEL system, the inflowing scattering cloud, and the X-ray properties (obscuration, scattering, FeK emission line, and variability) of this optically identified NLS2, are worth detailed follow-up observations. A detailed modeling of the NELR and scattering region of J1203+1624 based on their tight association is needed to interpret the interesting facts being observed. An interesting sub-class of AGNs with obscured SMBHs of a small mass accreting at a high rate is represented by J1203+1624. Their existence suggests that NLS1s follow the AGN unification scheme. In addition, the peculiar SED behavior, heavy extinction, and scattering in type 2 AGNs enables detailed exploration of both the active nucleus and the host galaxy. A sample of J1203+1624 analogues is important for understanding the SMBH-galaxy correlation and their coevolution for NLS1s, which will be conducted in our future work.

\acknowledgments
We appreciate the detailed and helpful comments from the referee, Robert R. J. Antonucci, which greatly improved the quality of this paper. We thank Victor Manuel Patino Alvarez for suggestions about English writing and some helpful comments. 
This work is supported by the National Natural Science Foundation of China (NSFC-11473025, 11573024, 11421303) 
and National Basic Research Program of China (the 973 Program 2013CB834905). 
HL, XP, and LS are supported by Natural Science Foundation of Anhui (1808085MA24).
PJ is supported by the National Natural Science Foundation of China
(NSFC-11233002). We acknowledge the use of the Multiple Mirror Telescope at Fred Lawrence Whipple Observatory
and the Hale 200-inch Telescope at Palomar Observatory through
the Telescope Access Program (TAP), as well as the archival data from the GALEX, SDSS, 2MASS and $WISE$ Surveys. TAP is funded by the Strategic Priority Research Program, The Emergence of Cosmological Structures
(XDB09000000), National Astronomical Observatories, Chinese Academy of Sciences, and the Special Fund
for Astronomy from the Ministry of Finance. Observations obtained with the Hale Telescope
 were obtained as part of an agreement between the National Astronomical Observatories, Chinese
Academy of Sciences, and the California Institute of Technology. Funding for SDSS-III has been provided
by the Alfred P. Sloan Foundation, the Participating Institutions, the National Science Foundation, and the
U.S. Department of Energy Office of Science. The SDSS-III website is http:// www.sdss3.org/.


\begin{thebibliography}{}
\bibitem[Abdo et al.(2009)]{2009ApJ...707L.142A} Abdo, A.~A., Ackermann, M., Ajello, M., et al.\ 2009, \apjl, 707, L142
\bibitem[Antonucci(1984)]{1984ApJ...278..499A} Antonucci, R.~R.~J.\ 1984, \apj, 278, 499
\bibitem[Antonucci \& Miller(1985)]{1985ApJ...297..621A} Antonucci, R.~R.~J., \& Miller, J.~S.\ 1985, \apj, 297, 621 
\bibitem[Antonucci et al.(1989)]{1989ApJ...342...64A} Antonucci, R.~R.~J., Kinney, A.~L., \& Ford, H.~C.\ 1989, \apj, 342, 64 
\bibitem[Antonucci(1993)]{1993ARA&A..31..473A} Antonucci, R.\ 1993, \araa, 31, 473
\bibitem[Baldwin et al.(1981)]{1981PASP...93....5B} Baldwin, J.~A., Phillips, M.~M., \& Terlevich, R.\ 1981, \pasp, 93, 5 
\bibitem[Baskin \& Laor(2005)]{2005MNRAS.358.1043B} Baskin, A., \& Laor, A.\ 2005, \mnras, 358, 1043 
\bibitem[Bruzual \& Charlot (2003)]{} Bruzual, G.\ \& Charlot, S.\ 2003, MNRAS, 344, 1000
\bibitem[Dewangan \& Griffiths(2005)]{2005ApJ...625L..31D} Dewangan, G.~C., \& Griffiths, R.~E.\ 2005, \apjl, 625, L31
\bibitem[Fitzpatrick \& Massa(2007)]{2007ApJ...663..320F} Fitzpatrick, E.~L., \& Massa, D.\ 2007, \apj, 663, 320
\bibitem[Gelbord et al.(2009)]{2009MNRAS.397..172G} Gelbord, J.~M., Mullaney, J.~R., \& Ward, M.~J.\ 2009, \mnras, 397, 172
\bibitem[Goodrich \& Miller(1988)]{1988ApJ...331..332G} Goodrich, R.~W., \& Miller, J.~S.\ 1988, \apj, 331, 332 
\bibitem[Goodrich(1989)]{1989ApJ...342..224G} Goodrich, R.~W.\ 1989, \apj, 342, 224

\bibitem[Greene \& Ho(2005)]{2005ApJ...630..122G} Greene, J.~E., \& Ho, L.~C.\ 2005, \apj, 630, 122
\bibitem[Grier et al.(2013)]{2013ApJ...764...47G} Grier, C.~J., Peterson, B.~M., Horne, K., et al.\ 2013, \apj, 764, 47 

\bibitem[Jiang et al.(2013)]{2013AJ....145..157J} Jiang, P., Zhou, H., Ji, T., et al.\ 2013, \aj, 145, 157
\bibitem[Kauffmann et al.(2003)]{2003MNRAS.346.1055K} Kauffmann, G., Heckman, T.~M., Tremonti, C., et al.\ 2003, \mnras, 346, 1055 

\bibitem[Kewley et al.(2001)]{2001ApJ...556..121K} Kewley, L.~J., Dopita, M.~A., Sutherland, R.~S., Heisler, C.~A., \& Trevena, J.\ 2001, \apj, 556, 121 
\bibitem[Komossa et al.(2006)]{2006AJ....132..531K} Komossa, S., Voges, W., Xu, D., et al.\ 2006, \aj, 132, 531
\bibitem[Komossa(2008)]{2008RMxAC..32...86K} Komossa, S.\ 2008, Revista Mexicana de Astronomia y Astrofisica Conference Series, 32, 86
\bibitem[Leighly(1999a)]{1999ApJS..125..297L} Leighly, K.~M.\ 1999, \apjs, 125, 297
\bibitem[Leighly(1999b)]{1999ApJS..125..317L} Leighly, K.~M.\ 1999, \apjs, 125, 317

\bibitem[Li et al.(2013)]{2013ApJ...779..110L} Li, Y.-R., Wang, J.-M., Ho, L.~C., Du, P., \& Bai, J.-M.\ 2013, \apj, 779, 110 

\bibitem[Li et al.(2015)]{2015ApJ...812...99L} Li, Z., Zhou, H., Hao, L., et al.\ 2015, \apj, 812, 99
\bibitem[Liu et al.(2000)]{2000MNRAS.312..585L} Liu, X.-W., Storey, P.~J., Barlow, M.~J., et al.\ 2000, \mnras, 312, 585 


\bibitem[Martin et al.(1989)]{1989A&A...215..219M} Martin, N., Maurice, E., \& Lequeux, J.\ 1989, \aap, 215, 219
\bibitem[Mathis et al.(1977)]{1977ApJ...217..425M} Mathis, J.~S., Rumpl, W., \& Nordsieck, K.~H.\ 1977, \apj, 217, 425 

\bibitem[Miller \& Goodrich(1990)]{1990ApJ...355..456M} Miller, J.~S., \& Goodrich, R.~W.\ 1990, \apj, 355, 456 
\bibitem[Miller et al.(1991)]{1991ApJ...378...47M} Miller, J.~S., Goodrich, R.~W., \& Mathews, W.~G.\ 1991, \apj, 378, 47 % Spectropolarimetry of NGC1068
\bibitem[Morrissey et al.(2007)]{2007ApJS..173..682M} Morrissey, P., Conrow, T., Barlow, T.~A., et al.\ 2007, \apjs, 173, 682
\bibitem[Nagar et al.(2002)]{2002A&A...391L..21N} Nagar, N.~M., Oliva, E., Marconi, A., \& Maiolino, R.\ 2002, \aap, 391, L21
%\bibitem[Nagao et al.(2004)]{2004AJ....128.2066N} Nagao, T., Kawabata, K.~S., Murayama, T., et al.\ 2004, \aj, 128, 2066
% \bibitem[Nenkova et al.(2008)]{2008ApJ...685..160N} Nenkova, M., Sirocky, M.~M., Nikutta, R., Ivezi{\'c}, {\v Z}., \& Elitzur, M.\ 2008, \apj, 685, 160 

\bibitem[Osterbrock(1981)]{1981ApJ...249..462O} Osterbrock, D.~E.\ 1981, \apj, 249, 462
\bibitem[Osterbrock \& Ferland(2006)]{2006agna.book.....O} Osterbrock, D.~E., \& Ferland, G.~J.\ 2006, Astrophysics of gaseous nebulae and active galactic nuclei, 2nd.~ed.~by D.E.~Osterbrock and G.J.~Ferland.~Sausalito, CA: University Science Books, 2006,
\bibitem[Pennell et al.(2017)]{2017MNRAS.468.1433P} Pennell, A., Runnoe, J.~C., \& Brotherton, M.~S.\ 2017, \mnras, 468, 1433 
\bibitem[Peterson(1997)]{1997iagn.book.....P} Peterson, B.~M.\ 1997, An introduction to active galactic nuclei, Publisher: Cambridge, New York Cambridge University Press, 1997 Physical description xvi, 238 p.~ISBN 0521473489,  
\bibitem[Pogge(2000)]{2000NewAR..44..381P} Pogge, R.~W.\ 2000, NewAR, 44, 381

%\bibitem[Rakshit et al.(2017)]{2017ApJS..229...39R} Rakshit, S., Stalin, C.~S., Chand, H., \& Zhang, X.-G.\ 2017, \apjs, 229, 39

\bibitem[Ramos Almeida et al.(2008)]{2008ApJ...680L..17R} Ramos Almeida, C., P{\'e}rez Garc{\'{\i}}a, A.~M., Acosta-Pulido, J.~A., \& Gonz{\'a}lez-Mart{\'{\i}}n, O.\ 2008, \apjl, 680, L17
\bibitem[Ramos Almeida et al.(2016)]{2016MNRAS.461.1387R} Ramos Almeida, C., Mart{\'{\i}}nez Gonz{\'a}lez, M.~J., Asensio Ramos, A., et al.\ 2016, \mnras, 461, 1387
\bibitem[Rashed et al.(2015)]{2015MNRAS.454.2918R} Rashed, Y.~E., Eckart, A., Valencia-S., M., et al.\ 2015, \mnras, 454, 2918 
\bibitem[Robinson et al.(2011)]{2011ASPC..449..431R} Robinson, A., Young, S., Axon, D.~J., \& Smith, J.~E.\ 2011, Astronomical Polarimetry 2008: Science from Small to Large Telescopes, 449, 431 

\bibitem[Rokaki et al.(2003)]{2003MNRAS.340.1298R} Rokaki, E., Lawrence, A., Economou, F., \& Mastichiadis, A.\ 2003, \mnras, 340, 1298
 \bibitem[Rose et al.(2015)]{2015MNRAS.448.2900R} Rose, M., Elvis, M., \& Tadhunter, C.~N.\ 2015, \mnras, 448, 2900  
\bibitem[Schlafly \& Finkbeiner(2011)] Schlafly, E. F., \& Finkbeiner, D. P. 2011, \apj, 737, 103
\bibitem[Schmidt et al.(1989)]{1989PASP..101..713S} Schmidt, G.~D., Weymann, R.~J., \& Foltz, C.~B.\ 1989, \pasp, 101, 713
\bibitem[Schmidt et al.(1992)]{1992ApJ...398L..57S} Schmidt, G.~D., Stockman, H.~S., \& Smith, P.~S.\ 1992, \apjl, 398, L57
%\bibitem[Shen \& Ho(2014)]{2014Natur.513..210S} Shen, Y., \& Ho, L.~C.\ 2014, \nat, 513, 210 
\bibitem[Skrutskie et al.(2006)]{2006AJ....131.1163S} Skrutskie, M.~F., Cutri, R.~M., Stiening, R., et al.\ 2006, \aj, 131, 1163
\bibitem[Smith et al.(2000)]{2000ApJ...545L..19S} Smith, P.~S., Schmidt, G.~D., Hines, D.~C., Cutri, R.~M., \& Nelson, B.~O.\ 2000, \apjl, 545, L19 
\bibitem[Smith et al.(2003)]{2003ApJ...593..676S} Smith, P.~S., Schmidt, G.~D., Hines, D.~C., \& Foltz, C.~B.\ 2003, \apj, 593, 676
\bibitem[Tarchi et al.(2011)]{2011A&A...532A.125T} Tarchi, A., Castangia, P., Columbano, A., Panessa, F., \& Braatz, J.~A.\ 2011, \aap, 532, A125 
\bibitem[Tran et al.(1992)]{1992ApJ...397..452T} Tran, H.~D., Miller, J.~S., \& Kay, L.~E.\ 1992, \apj, 397, 452
\bibitem[Tremaine et al.(2002)]{2002ApJ...574..740T} Tremaine, S., Gebhardt, K., Bender, R., et al.\ 2002, \apj, 574, 740
\bibitem[Vanden Berk et al.(2001)]{2001AJ....122..549V} Vanden Berk, D.~E.,
Richards, G.~T., Bauer, A., et al.\ 2001, \aj, 122, 549

\bibitem[V{\'e}ron-Cetty et al.(2001)]{2001A&A...372..730V} V{\'e}ron-Cetty, M.-P., V{\'e}ron, P., \& Gon{\c c}alves, A.~C.\ 2001, \aap, 372, 730
\bibitem[V{\'e}ron-Cetty et al.(2004)]{2004A&A...417..515V} V{\'e}ron-Cetty, M.-P., Joly, M., \& V{\'e}ron, P.\ 2004, \aap, 417, 515

\bibitem[Voges et al.(1999)]{1999A&A...349..389V} Voges, W., Aschenbach, B., Boller, T., et al.\ 1999, \aap, 349, 389
\bibitem[Wills \& Browne(1986)]{1986ApJ...302...56W} Wills, B.~J., \& Browne, I.~W.~A.\ 1986, \apj, 302, 56 

%\bibitem[Williams et al.(2002)]{william02} Williams, R.~J., Pogge, R.~W., \& Mathur, S.\ 2002, \aj, 124, 3042
\bibitem[Wright et al.(2010)]{2010AJ....140.1868W} Wright, E.~L., Eisenhardt, P.~R.~M., Mainzer, A.~K., et al.\ 2010, \aj, 140, 1868
\bibitem[York et al.(2000)]{2000AJ....120.1579Y} York, D.~G., Adelman, J., Anderson, J.~E., Jr., et al.\ 2000, \aj, 120, 1579
%\bibitem[Young et al.(2007)]{2007Natur.450...74Y} Young, S., Axon, D.~J., Robinson, A., Hough, J.~H., \& Smith, J.~E.\ 2007, \nat, 450, 74 
\bibitem[Yuan et al.(2008)]{2008ApJ...685..801Y} Yuan, W., Zhou, H.~Y., Komossa, S., et al.\ 2008, \apj, 685, 801-827
\bibitem[Zafar et al.(2015)]{2015A&A...584A.100Z} Zafar, T., M{\o}ller, P., Watson, D., et al.\ 2015, \aap, 584, A100 

\bibitem[Zakamska et al.(2003)]{2003AJ....126.2125Z} Zakamska, N.~L., Strauss, M.~A., Krolik, J.~H., et al.\ 2003, \aj, 126, 2125
\bibitem[Zakamska et al.(2005)]{2005AJ....129.1212Z} Zakamska, N.~L., Schmidt, G.~D., Smith, P.~S., et al.\ 2005, \aj, 129, 1212 
\bibitem[Zhang et al.(2017a)]{2017ApJ...836...86Z} Zhang, S., Zhou, H., Shi, X., et al.\ 2017, \apj, 836, 86

\bibitem[Zhang et al.(2017b)]{2017ApJ...845..126Z} Zhang, S., Zhou, H., Shi, X., et al.\ 2017, \apj, 845, 126
\bibitem[Zhou et al.(2002)]{2002ApJ...581...96Z} Zhou, H.-Y., Wang, T.-G., Zhou, Y.-Y., Cheng-Li, \& Dong, X.-B.\ 2002, \apj, 581, 96 
\bibitem[Zhou et al.(2006)]{zhou06} Zhou, H., Wang, T., Yuan, W., et al.\ 2006, \apjs, 166, 128
\bibitem[Zhou et al.(2010)]{2010ApJ...708..742Z} Zhou, H., Ge, J., Lu, H., et al.\ 2010, \apj, 708, 742

\end{thebibliography}
\end{document}